\documentclass[12pt]{article}
\usepackage{graphicx}
\usepackage{lscape}
\usepackage{citesort}
\usepackage{amssymb}
\usepackage{appendix}
\usepackage{multirow}
\usepackage[table]{xcolor}
\usepackage{colortbl}
\definecolor{lightgray}{gray}{0.9}

\usepackage{mathrsfs}
\begin{document}
\def\qq{\langle \bar q q \rangle}
\def\uu{\langle \bar u u \rangle}
\def\dd{\langle \bar d d \rangle}
\def\sp{\langle \bar s s \rangle}
\def\GG{\langle g_s^2 G^2 \rangle}
\def\Tr{\mbox{Tr}}
\def\figt#1#2#3{
        \begin{figure}
        $\left. \right.$
        \vspace*{-2cm}
        \begin{center}
        \includegraphics[width=10cm]{#1}
        \end{center}
        \vspace*{-0.2cm}
        \caption{#3}
        \label{#2}
        \end{figure}
    }

\def\figb#1#2#3{
        \begin{figure}
        $\left. \right.$
        \vspace*{-1cm}
        \begin{center}
        \includegraphics[width=10cm]{#1}
        \end{center}
        \vspace*{-0.2cm}
        \caption{#3}
        \label{#2}
        \end{figure}
                }

\def\ds{\displaystyle}
\def\beq{\begin{equation}}
\def\eeq{\end{equation}}
\def\bea{\begin{eqnarray}}
\def\eea{\end{eqnarray}}
\def\beeq{\begin{eqnarray}}
\def\eeeq{\end{eqnarray}}
\def\ve{\vert}
\def\vel{\left|}
\def\ver{\right|}
\def\nnb{\nonumber}
\def\ga{\left(}
\def\dr{\right)}
\def\aga{\left\{}
\def\adr{\right\}}
\def\lla{\left<}
\def\rra{\right>}
\def\rar{\rightarrow}
\def\lrar{\leftrightarrow}
\def\nnb{\nonumber}
\def\la{\langle}
\def\ra{\rangle}
\def\ba{\begin{array}}
\def\ea{\end{array}}
\def\tr{\mbox{Tr}}
\def\ssp{{\Sigma^{*+}}}
\def\sso{{\Sigma^{*0}}}
\def\ssm{{\Sigma^{*-}}}
\def\xis0{{\Xi^{*0}}}
\def\xism{{\Xi^{*-}}}
\def\qs{\la \bar s s \ra}
\def\qu{\la \bar u u \ra}
\def\qd{\la \bar d d \ra}
\def\qq{\la \bar q q \ra}
\def\gGgG{\la g^2 G^2 \ra}
\def\q{\gamma_5 \not\!q}
\def\x{\gamma_5 \not\!x}
\def\g5{\gamma_5}
\def\sb{S_Q^{cf}}
\def\sd{S_d^{be}}
\def\su{S_u^{ad}}
\def\sbp{{S}_Q^{'cf}}
\def\sdp{{S}_d^{'be}}
\def\sup{{S}_u^{'ad}}
\def\ssp{{S}_s^{'??}}

\def\sig{\sigma_{\mu \nu} \gamma_5 p^\mu q^\nu}
\def\fo{f_0(\frac{s_0}{M^2})}
\def\ffi{f_1(\frac{s_0}{M^2})}
\def\fii{f_2(\frac{s_0}{M^2})}
\def\O{{\cal O}}
\def\sl{{\Sigma^0 \Lambda}}
\def\es{\!\!\! &=& \!\!\!}
\def\ap{\!\!\! &\approx& \!\!\!}
\def\md{\!\!\!\! &\mid& \!\!\!\!}
\def\ar{&+& \!\!\!}
\def\ek{&-& \!\!\!}
\def\kek{\!\!\!&-& \!\!\!}
\def\cp{&\times& \!\!\!}
\def\se{\!\!\! &\simeq& \!\!\!}
\def\eqv{&\equiv& \!\!\!}
\def\kpm{&\pm& \!\!\!}
\def\kmp{&\mp& \!\!\!}
\def\mcdot{\!\cdot\!}
\def\erar{&\rightarrow&}
\def\olra{\stackrel{\leftrightarrow}}
\def\ola{\stackrel{\leftarrow}}
\def\ora{\stackrel{\rightarrow}}

\def\simlt{\stackrel{<}{{}_\sim}}
\def\simgt{\stackrel{>}{{}_\sim}}


\title{
         {\Large
                 {\bf
                     Properties of nucleon in nuclear matter: once more
                 }
         }
      }

\author{\vspace{1cm}\\
{\small  K. Azizi$^1$ \thanks {e-mail: kazizi@dogus.edu.tr}\,\,, N. Er$^2$ \thanks {e-mail: nuray@ibu.edu.tr}}  \\
{\small $^1$ Department of Physics, Do\u gu\c s University,
Ac{\i}badem-Kad{\i}k\"oy, 34722 \.{I}stanbul, Turkey}\\
{\small $^2$ Department of Physics, Abant {I}zzet Baysal University,
G\"olk\"oy Kamp\"us\"u, 14980, Bolu, Turkey}\\
}
\date{}

\begin{titlepage}
\maketitle
\thispagestyle{empty}
\begin{abstract} 
We calculate the mass and residue of the nucleon in nuclear matter in the frame
work of QCD sum rules using the   nucleon's interpolating
current with an arbitrary mixing parameter. We evaluate the effects of the nuclear medium on these quantities
and compare the obtained results with the existing theoretical predictions. The results are also
compared with those obtained in vacuum to find the shifts in the quantities
under consideration. Our calculations show that these shifts in the mass
and residue are about $32\%$ and $15\%$, respectively. 

PACS number(s):11.55.Hx, 21.65.-f, 14.20.-c, 14.20.Dh
\end{abstract}
\end{titlepage}


\section{Introduction}  

To analyze the experimental results on the relativistic heavy ion collision
held at different experiments such as CERN, the European Organization for
Nuclear Research, and BNL, Brookhaven National Laboratory, as well as for better
understanding the internal structure of the neutron stars,
the in-medium properties of hadrons especially the properties of nucleons at
nuclear medium are needed. From the experimental side, there has been a good
progress on the in-medium properties of hadrons in recent years. The FAIR (Facility for Antiproton and Ion Research), and CBM (the Compressed
Baryonic Matter) Collaboration at GSI intend to study the in-medium effects on
the parameters of different hadrons. The Panda Collaboration, on the other hand,
aims to concentrate on the properties of the charmed hadrons and study  probable shifts on
their masses and widths in nuclear medium \cite{Lutz,Friman}.

From  theoretical side, there are dozens of works devoted to the study of
the nuclear matter and properties of hadrons especially nucleons at dense medium.
In \cite{Drukarev1}, the basic properties of the  nuclear matter are determined
in the frame work of QCD sum rules as one of the most applicable and attractive
tools to hadron physics. This method, then, has been applied to some
finite-density problems \cite{DrukarevA,hatsuda,adami}.
In \cite{DrukarevA}, the authors have used the
finite density sum rules to investigate the saturation properties of nuclear
matter. In series of papers \cite{RJF,XJ1,Nielsen}, T. D. Cohen et al, have applied the QCD sum
rules to relativistic nuclear physics and studied the effects of nuclear matter
on the mass of the nucleons mostly for the Ioffe current. Only in \cite{Nielsen}, the
authors a bit extend the Ioffe current ($\beta=-1$) with $\beta$ being a mixing
parameter in the interpolating current of the nucleons to $-1.15 \leq \beta
\leq -0.95$ using the mass sum rules. For some studies of nucleon mass shift in nuclear medium for the Ioffe 
current and properties of other hadrons in dense medium see for instance \cite{TDC2,Cohen45,Cohen,Drukarev,Hogaasen,Wang2011,Wang2012,Wang,Lee,Asakawa,Leinweber,Klingl,Leupold,Hayashigaki,EGDur,Hilger,Yasui}. 
The effects of four-quark condensate on the  nucleon parameters have also been studied in \cite{thomas}.
Recently, the QCD sum rules has been used to analyze the residue of the nucleon pole 
as a function of nuclear density \cite{Mallik} using a special current corresponding to an axial-vector
diquark coupled to a quark. Note that the mass and residue of nucleon have also  been investigated in instanton medium very recently in \cite{drukarev2013}.  

In this article, we extend the previous studies to calculate both the mass and residue of the nucleon in nuclear matter using the 
interpolating current with an arbitrary mixing parameter  in the frame work of QCD sum rules. As the mass sum rule is the ratio of two sum rules (according to the method used), 
 it may not lead to a reliable region for the arbitrary mixing parameter. Since  the unstable points of two sum rules in nominator and denominator generally coincide and cancel each other and the mass shows
roughly a good stability with respect to the mixing parameter in the whole interval ($-\infty$,$+\infty$) for $\beta$. Hence, to restrict this parameter, the  only reliable chance is to use the sum rule for the residue
 as it does not contain any  ratio of sum rules and includes only one function from the operator product expansion (OPE) side.  In this connection,  we use the sum rule for the residue to find the 
working region for the mixing parameter $\beta$.  By this way, we  extend the previous calculations \cite{Mallik} on the residue as our working region for $\beta$ includes the Ioffe current used in \cite{Mallik}
to discuss the behavior of the  residue of the nucleon with respect to the nuclear density. We also extend the study \cite{Nielsen} on the mass of the nucleon in nuclear matter by the extension of the working region for the mixing parameter.
 Using the obtained working region for
 $\beta$ as well as working regions of other auxiliary parameters entered the sum rules, 
 we then obtain the shifts in the values of mass and residue compared to their vacuum values. We also compare our results on the mass and residue of the
nucleon with the existing numerical results 
obtained via Ioffe current in vacuum. Finally, we extract the vector and scalar self-energies of the nucleon in nuclear matter and compare the obtained results with the  predictions of
  model independent studies \cite{plohl1,plohl2}.

Our results on the mass and residue can be used  in theoretical 
calculations via the  current under consideration such as computation of the electromagnetic properties and multiple moments of the nucleon and 
the strong coupling constants of the nucleon to other hadrons in nuclear medium. Study of the electromagnetic properties of nucleons have been in the focus of many experimental and theoretical works
 for many years. Unfortunately, there is no a good consistency  among the obtained  results via different ways  on the electromagnetic form factors (see for instance \cite{ozpineci}  and references therein).
In  \cite{ozpineci} it is shown that  the Ioffe current fails 
to reproduce some experimental data on some of electromagnetic form factors of the nucleon in vacuum. 
Also, many  previous works on the strong coupling constants
 among various groups of baryons including the nucleons with different mesonic groups  (see for instance \cite{ozpineci2,savci}) reveal that the Ioffe current remains out of reliable  region 
and some other values of the mixing parameter $\beta$ for octet baryons are favored. We guess the problem of failing Ioffe current 
to explain some electromagnetic and strong coupling parameters in vacuum will still the case in also nuclear medium. When we calculate the electromagnetic or strong parameters of nucleon in nuclear matter
 using the current with an arbitrary mixing parameter,
 we immediately need their masses and residues calculated via the same current in nuclear matter. Our working region for the parameter $\beta$ as well as our predictions on the mass and residue of
the nucleon can be useful in this respect. 

The article is organized as follows: In section 2, we obtain  QCD sum rules for the mass and residue of the nucleon in the 
nuclear matter. Section 3 is devoted to the numerical analyses of the sum rules and our comparison of the results with the existing predictions.
Section 4 contains our concluding remarks.
\section{ QCD sum rules for the mass and residue of nucleon in nuclear matter}
To obtain the sum rules for the mass and residue of nucleon in nuclear matter, the
starting point is to consider the following two-point correlation function:
\begin{equation}\label{correilk}
\Pi(p)=i\int{d^4 xe^{ip\cdot x}\langle\psi_0|T[J
(x)\bar{J}(0)]|\psi_0\rangle},
\end{equation}
where $p$ is the four momentum of the nucleon and $|\psi_0\rangle$ is the nuclear matter ground
state. The   nucleon interpolating current is taken as 
\begin{equation}
J(x)=2\epsilon_{abc}\Sigma_{i=1}^{2}\Bigg[q_{1}^{T,a}(x)CA_{1}^{i}q_{2}^{b}(x)
\Bigg]A_{2}^{i}q_{1}^{c}(x),
\end{equation}
where $a, b, c$ are color indices, $C$ is the charge conjugation operator and
$A_{1}^{1}=I$, $A_{1}^{2}=A_{2}^{1}=\gamma_5$, 
$A_{2}^{2}=\beta$. As previously said, the parameter $\beta$ is an arbitrary auxiliary parameter,
and $\beta=-1$ corresponds to the Ioffe current (for some discussions about the nucleon interpolating currents see for instance \cite{drukarev2013,thomas,leinweber,stein}). The quark flavors for the proton (neutron) are $q_1=u$ and
$q_2=d$ ($q_1=d$ and $q_2=u$). Here we will adopt the isospin symmetry to treat the proton and neutron as 
nucleon.

From the general philosophy of the method under consideration,  we calculate the above mentioned correlation function from 
two different windows: in  terms of hadronic parameters called as the phenomenological or hadronic side and in terms of QCD
degrees of freedom using the OPE at nuclear medium named as OPE or theoretical side. Equating these two sides, 
we gain QCD sum rules for the mass and residue in nuclear matter. 
To suppress contribution of the higher states and continuum Borel transformation and continuum subtraction \cite{shif,col} are applied to both sides 
of the obtained sum rules. 

\subsection{Hadronic side}
In hadronic side, the correlation function is calculated inserting a complete set of nucleon state with the same quantum numbers as the interpolating current. 
After performing integral over four-x, we get
\begin{equation}\label{corre}
\Pi^{Had}(p)=-\frac{\langle\psi_0|J
(x)|N(p,s)\rangle\langle N(p,s)|\bar{J}(0)|\psi_0\rangle}{p^2-m_{N}^{*2}}+... ,
\end{equation}
where dots represents the contributions of higher states and continuum and $m_{N}^{*}$ is the modified mass of the nucleon in nuclear matter. The
matrix element of the interpolating current between the nucleon ground state
and the baryonic state is parametrized as
\begin{equation}\label{intcur}
\langle\psi_0|J(x)|N(p,s)\rangle=\lambda_{N}^{*}u(p,s) ,
\end{equation}
here $\lambda_{N}^{*}$ is the modified residue or the coupling strength of the nucleon current
$J(x)$ to the nucleon quasi-particle in the nuclear matter and $u(p,s)$ is
their positive energy Dirac spinor. Using  Eq. (\ref{intcur}) in Eq. (\ref{corre}), we get 
\begin{equation}
\Pi^{Had}(p)=-\frac{\lambda_{N}^{*2}(\!\not\!{p}+m_{N}^{*})}{p^2-m_{N}^{*2}}+... =-\frac{\lambda_{N}^{*2}}{(\!\not\!{p}-m_{N}^{*})}+... .
\end{equation}
Considering the interactions between the
nucleon and the nuclear matter, the hadronic side of the correlation function
takes the modified form
\begin{equation}\label{corre1}
\Pi^{Had}(p)=-\frac{\lambda_{N}^{*2}}{
(p^\mu-\Sigma_\nu^\mu)\gamma_\mu-(m_N+\Sigma_S)}+...,
\end{equation}
where $\Sigma_\nu^\mu$ and $\Sigma_S$ are vector and scalar self-energies of the nucleon in nuclear matter, respectively \cite{TDC2}. 
In general, we can write 
\begin{equation}\label{sigma}
\Sigma_\nu^\mu=\Sigma_\nu u^\mu+\Sigma'_\nu p^\mu,
\end{equation}
where $\Sigma_\nu$ and $\Sigma'_\nu$ are constants and $u^\mu$ is the four velocity of the nuclear medium. Here we neglect $\Sigma'_\nu$ due to its small contribution  (see also \cite{TDC2}).
Apart from the vacuum QCD calculations, the four-velocity of the nuclear matter is
new concept that causes extra structures to the correlation function. We shall work in the
rest frame of the nucleon with $u^\mu=(1,0)$. Substitution Eq. (\ref{sigma}) into Eq. (\ref{corre1}), the hadronic side of the correlation function becomes
\begin{equation}
\Pi^{Had}(p)=-\frac{\lambda_{N}^{*2}}{
(\!\not\!{p}-\Sigma_\nu\!\not\!{u})-(m_N+\Sigma_S)}+... ,
\end{equation} 
which can be written in terms of three different structure as 
\begin{equation}
\Pi^{Had}(p)=\Pi^{Had}_p(p^2,p_0)\!\not\!{p}+\Pi^{Had}_u(p^2,p_0)\!\not\!{u}
+\Pi^{Had}_S(p^2,p_0)I,
\end{equation}
where $p_0$ is the energy of the quasi-particle, $I$ is the unit matrix and
\begin{eqnarray}
\Pi^{Had}_p(p^2,p_0)&=&-\lambda_N^{*2}\frac{1}{p^2-\mu^2},\nonumber \\
\Pi^{Had}_u(p^2,p_0)&=&+\lambda_N^{*2}\frac{\Sigma_\nu}{p^2-\mu^2},
\nonumber \\
\Pi^{Had}_S(p^2,p_0)&=&-\lambda_N^{*2}\frac{m_N^*}{p^2-\mu^2} .
\end{eqnarray}
Here $m_N^*=m_N+\Sigma_S$ and
$\mu^2=m_N^{*2}-\Sigma_{\nu}^2+2p_0\Sigma_\nu$. After a Wick rotation and applying the Borel transformation with respect to $p^2$, we get 
\begin{eqnarray}
\hat{B}\Pi^{Had}_p(p^2,p_0)&=&-\lambda_N^{*2}e^{-\mu^2/M^2}, \nonumber \\
\hat{B}\Pi^{Had}_u(p^2,p_0)&=&+\lambda_N^{*2}\Sigma_\nu e^{-\mu^2/M^2},
\nonumber \\
\hat{B}\Pi^{Had}_S(p^2,p_0)&=&-\lambda_N^{*2}m_N^* e^{-\mu^2/M^2},
\end{eqnarray}
where $M^2$ is the Borel mass parameter, we shall find its working region in Section 3. 

\subsection{OPE side}
The OPE side of the correlation function can be calculated  in deep Euclidean region. This function can also be 
written in terms of different structures  as 
\begin{equation}
\Pi^{OPE}(p)=\Pi_{p}^{OPE}\!\not\! {p}+\Pi_{u}^{OPE}\!\not\!
{u}+\Pi_{S}^{OPE}I.
\end{equation}
Each $\Pi_{i}^{OPE}$ function, where $i=\!\not\!{p}, \!\not\!{u}$ and $I$, can be written in terms of dispersion integral as  
\begin{equation}
\Pi^{OPE}_{i}=\int\frac{\rho_{i}(s)}{s-p^{2}}ds,
\end{equation}
where $\rho_{i}(s)=\frac{1}{\pi}\textrm{Im}[\Pi^{OPE}_{i}]$ are the spectral densities. 
Using the explicit form of the interpolating current in correlation function of  Eq. (\ref{correilk}) and contracting out
all quark pairs via Wick's theorem, we find
\begin{eqnarray}\label{corre2}
\Pi^{OPE}(p) &=&
-4i\epsilon_{abc}\epsilon_{a'b'c'}\int d^4 x e^{ipx}\Bigg\langle \psi_0\Bigg
|\Bigg\{\Bigg(\gamma_{5}S^{cb'}_{u}(x)S'^{ba'}_{d}(x)S^{ac'}_{u}(x)\gamma_{5}
\nonumber \\
&-&\gamma_{5}S^{cc'}_{u}(x)\gamma_{5}Tr\Bigg[S^{ab'}_{u}(x)S'^{ba'}_{d}(x)\Bigg]
\Bigg)
+\beta\Bigg(\gamma_{5}S^{cb'}_{u}(x)\gamma_{5}S'^{ba'}_{d}(x)S^{ac'}_{u}
(x)\nonumber \\
&+&S^{cb'}_{u}(x)S'^{ba'}_{d}(x)\gamma_{5}S^{ac'}_{u}(x)\gamma_{5}
-\gamma_{5}S^{cc'}_{u}(x)Tr\Bigg[S^{ab'}_{u}(x)\gamma_{5}S'^{ba'}_{d}(x)\Bigg]
\nonumber \\
&-&S^{cc'}_{u}(x)\gamma_{5}Tr\Bigg[S^{ab'}_{u}(x)S'^{ba'}_{d}(x)\gamma_{5}\Bigg]
\Bigg)
+\beta^2\Bigg(S^{cb'}_{u}(x)\gamma_{5}S'^{ba'}_{d}(x)\gamma_{5}S^{ac'}_{u}
(x)\nonumber \\
&-&S^{cc'}_{u}(x)Tr\Bigg[S^{ba'}_{d}(x)\gamma_{5}S'^{ab'}_{u}(x)\gamma_{5}\Bigg]
\Bigg)
\Bigg\}\Bigg| \psi_0\Bigg\rangle,
\end{eqnarray}
where $S'=CS^TC$, $S_{u,d}$ are light quarks propagators and $Tr[...]$ denotes trace of gamma matrices.  
In coordinate-space, the light quark
propagator at the nuclear medium has the following form in the fixed-point gauge \cite{Cohen,yasaki}:
\begin{eqnarray}\label{propagator}
 S_{q}^{ab}(x)&\equiv& \langle\psi_0|T[q^a
(x)\bar{q}^b(0)]|\psi_0\rangle_{\rho_N} \nonumber \\
&=&
\frac{i}{2\pi^2}\delta^{ab}\frac{1}{(x^2)^2}\not\!x
-\frac{m_q }{ 4\pi^2} \delta^ { ab } \frac { 1}{x^2} \nonumber \\
&+&
\chi^a_q(x)\bar{\chi}^b_q(0)-\frac{ig_s}{32\pi^2}F_{\mu\nu}^A(0)t^{ab,A
}\frac{1}{x^2}[\not\!x\sigma^{\mu\nu}+\sigma^{\mu\nu}\not\!x]+...,
\end{eqnarray}
where $\rho_N$ is the nuclear matter density, $\chi^a_q$ and $\bar{\chi}^b_q$ are the Grassmann background quark fields, $F_{\mu\nu}^A$ is classical background gluon field; and  the first and second terms are the expansion of the free quark propagator
to first order in the quark mass (perturbative part), and the third and forth
terms are the contributions due to the background quark and gluon fields
(non-perturbative part), respectively. The gluonic contribution to above equation corresponds to a single gluon intraction keeping only the leading term in the short-distance expansion of the gluon field.
We ignore from contributions of the derivatives of the gluon field tensor as well as additional gluon interactions in the expresion of the light-quark propagator (see also \cite{Cohen}).
   When using Eq. (\ref{propagator}) in Eq. (\ref{corre2}), we  will end up with the products of the Grassmann background quark fields and classical background gluon fields which correspond to 
the ground-state matrix elements of the corresponding quark and gluon operators  \cite{Cohen}
\begin{eqnarray}\label{}
&&\chi_{a\alpha}^{q}(x)\bar{\chi}_{b\beta}^{q}(0)=\langle
q_{a\alpha}(x)\bar{q}_{ b\beta}(0)\rangle_{\rho_N}, ~~~~~~~
F_{\kappa\lambda}^{A}F_{\mu\nu}^{B}=\langle
G_{\kappa\lambda}^{A}G_{\mu\nu}^{B}\rangle_{\rho_N}, \nonumber \\
&&\chi_{a\alpha}^{q}\bar{\chi}_{b\beta}^{q}F_{\mu\nu}^{A}=\langle
q_{a\alpha}\bar{q}_{ b\beta}G_{\mu\nu}^{A}\rangle_{\rho_N}, 
~~~~~~~~\chi_{a\alpha}^{q}\bar{\chi}_{b\beta}^{q}\chi_{c\gamma}^{q}\bar
{\chi}_{d\delta}^{q}=\langle
q_{a\alpha}\bar{q}_{b\beta}
q_{c\gamma}\bar{q}_{d\delta}\rangle_{\rho_N}.
\end{eqnarray}

To proceed, we need to define the quark
and gluon and mixed condensates in nuclear matter. The matrix element $\langle
q_{a\alpha}(x)\bar{q}_{b\beta}(0)\rangle_{\rho_N}$ is projected out as \cite{Cohen} 
\begin{eqnarray} \label{ }
\langle
q_{a\alpha}(x)\bar{q}_{b\beta}(0)\rangle_{\rho_N}&=&-\frac{\delta_{ab}}{12}\Bigg
[\Bigg(\langle\bar{q}q\rangle_{\rho_N}+x^{\mu}\langle\bar{q}D_{\mu}q\rangle_{
\rho_N}
+\frac{1}{2}x^{\mu}x^{\nu}\langle\bar{q}D_{\mu}D_{\nu}q\rangle_{\rho_N}
+...\Bigg)\delta_{\alpha\beta}\nonumber \\
&&+\Bigg(\langle\bar{q}\gamma_{\lambda}q\rangle_{\rho_N}+x^{\mu}\langle\bar{q}
\gamma_{\lambda}D_{\mu} q\rangle_{\rho_N}
+\frac{1}{2}x^{\mu}x^{\nu}\langle\bar{q}\gamma_{\lambda}D_{\mu}D_{\nu}
q\rangle_{\rho_N}
+...\Bigg)\gamma^{\lambda}_{\alpha\beta} \Bigg].\nonumber \\
\end{eqnarray}
The quark-gluon condensate in nuclear matter is written as 
\begin{eqnarray} \label{ }
\langle
g_{s}q_{a\alpha}\bar{q}_{b\beta}G_{\mu\nu}^{A}\rangle_{\rho_N}&=&-\frac{t_{
ab}^{A
}}{96}\Bigg\{\langle g_{s}\bar{q}\sigma\cdot
Gq\rangle_{\rho_N}\Bigg[\sigma_{\mu\nu}+i(u_{\mu}\gamma_{\nu}-u_{\nu}\gamma_{\mu
})
\!\not\! {u}\Bigg]_{\alpha\beta} \nonumber \\
&&+\langle g_{s}\bar{q}\!\not\! {u}\sigma\cdot
Gq\rangle_{\rho_N}\Bigg[\sigma_{\mu\nu}\!\not\!
{u}+i(u_{\mu}\gamma_{\nu}-u_{\nu}\gamma_{\mu}
)\Bigg]_{\alpha\beta} \nonumber \\
&&-4\Bigg(\langle\bar{q}u\cdot D u\cdot D q\rangle_{\rho_N}+im_{q}\langle\bar{q}
\!\not\! {u}u\cdot D q\rangle_{\rho_N}\Bigg) \nonumber \\
&&\times\Bigg[\sigma_{\mu\nu}+2i(u_{\mu}\gamma_{\nu}-u_{\nu}\gamma_{\mu}
)\!\not\! {u}\Bigg]_{\alpha\beta}\Bigg\},
\end{eqnarray}
where $t_{ab}^{A}$ are Gell-Mann matrices and $D_\mu=\frac{1}{2}(\gamma_\mu \!\not\!{D}+\!\not\!{D}\gamma_\mu)$.
The matrix element of the four-dimension gluon condensate can also written as
\begin{equation}
 \langle
G_{\kappa\lambda}^{A}G_{\mu\nu}^{B}\rangle_{\rho_N}=\frac{\delta^{AB}}{96}
\Bigg[
\langle
G^{2}\rangle_{\rho_N}(g_{\kappa\mu}g_{\lambda\nu}-g_{\kappa\nu}g_{\lambda\mu}
)+O(\langle
\textbf{E}^{2}+\textbf{B}^{2}\rangle_{\rho_N})\Bigg],
\end{equation}
where we neglect the last term in this equation because of its small contribution.
The various condensates in the above equations are defined as \cite{Cohen,XJ1}
\begin{eqnarray} \label{ }
\langle\bar{q}\gamma_{\mu}q\rangle_{\rho_N}&=&\langle\bar{q}\!\not\!{u}q\rangle_
{\rho_N}
u_{\mu} , \\
\langle\bar{q}D_{\mu}q\rangle_{\rho_N}&=&\langle\bar{q}u\cdot D
q\rangle_{\rho_N}
u_{\mu}=-im_{q}\langle\bar{q}\!\not\!{u}q\rangle_{\rho_N}
u_{\mu}  ,\\
\langle\bar{q}\gamma_{\mu}D_{\nu}q\rangle_{\rho_N}&=&\frac{4}{3}\langle\bar{q}
\!\not\! {u}u\cdot D q\rangle_{\rho_N}(u_{\mu}u_{\nu}-\frac{1}{4}g_{\mu\nu}) 
+\frac{i}{3}m_{q} \langle\bar{q}q\rangle_{\rho_N}(u_{\mu}u_{\nu}-g_{\mu\nu}),
\\
\langle\bar{q}D_{\mu}D_{\nu}q\rangle_{\rho_N}&=&\frac{4}{3}\langle\bar{q}
u\cdot D u\cdot D q\rangle_{\rho_N}(u_{\mu}u_{\nu}-\frac{1}{4}g_{\mu\nu}) 
-\frac{1}{6} \langle
g_{s}\bar{q}\sigma\cdot Gq\rangle_{\rho_N}(u_{\mu}u_{\nu}-g_{\mu\nu}) ,\\
\langle\bar{q}\gamma_{\lambda}D_{\mu}D_{\nu}q\rangle_{\rho_N}&=&2\langle\bar{q}
\!\not\! {u}u\cdot D u\cdot D
q\rangle_{\rho_N}\Bigg[u_{\lambda}u_{\mu}u_{\nu} -\frac{1}{6} 
(u_{\lambda}g_{\mu\nu}+u_{\mu}g_{\lambda\nu}+u_{\nu}g_{\lambda\mu})\Bigg]\nonumber\\
&&-\frac{1}{6} \langle
g_{s}\bar{q}\!\not\! {u}\sigma\cdot
Gq\rangle_{\rho_N}(u_{\lambda}u_{\mu}u_{\nu}-u_{\lambda}g_{\mu\nu}),
\end{eqnarray}
where the  equations of motion have been used and $\textit{O}(m^2_q)$ terms have been neglected due to their very small contributions \cite{Cohen}.
Now, we use  the expressions of the light quark propagator in nuclear medium and  different condensates presented above in Eq. (\ref{corre2}) and perform the four-integral over $x$ to go to the momentum space.
 To suppress
the contributions of the higher states and continuum we apply  the Borel transformation with respect to the momentum squared  and perform continuum subtraction. We also 
 use the quark-hadron duality assumption. 
The  $\Pi_{i}^{OPE}$ functions can be written interms of the even and odd parts in terms of  $p_0$ as
\begin{eqnarray}
 \Pi_{i}^{OPE}=\Pi_{i}^{E}+p_0 \Pi_{i}^{O},
\end{eqnarray}
where, after lengthy calculations, for the  invariant functions $\Pi_{i}^{E,O}$  in Borel scheme we get
\begin{eqnarray}
 \hat{B}\Pi_{p}^{E}&=&-\frac{1}{256 \pi^{4}}\int_{0}^{s_{0}}ds
e^{-s/M^{2}}s^{2}\Bigg[5+\beta(2+5\beta)\Bigg]\nonumber \\
&&+\frac{1}{72 \pi^{2}}\int_{0}^{s_{0}}ds
e^{-s/M^{2}}\Bigg\{-8\Bigg[5+\beta(2+5\beta)\Bigg]m_{q}\langle
\bar{q}q\rangle_{\rho_N}\nonumber \\
&&+9(-1+\beta)\Bigg[3(1+\beta)m_{d}+2m_{u}+4\beta
m_{u}\Bigg]\langle
\bar{q}q\rangle_{\rho_N}\nonumber \\
&&+5\Big[5+\beta(2+5\beta)\bigg]\langle
q^{\dag}iD_{0}q\rangle_{\rho_N} 
\Bigg\}\nonumber \\
&&-\frac{\langle
g_{s}^{2}G^{2}\rangle_{\rho_N}}{1024 \pi^{4}}\int_{0}^{s_{0}}ds
e^{-s/M^{2}}(6+\beta+5\beta^{2}) \nonumber \\
&&+\frac{1}{192M^{2}\pi^{2}}\Bigg\{(-1+\beta)\Bigg[-\Big(40(1+\beta)m_{d}
+(26+43\beta)m_{u}\Big)M^2\nonumber \\
&&+8\Big(3(1+\beta)m_{d}+2m_{u}+4\beta
m_{u}\Big)p_0^{2}\Bigg] \Bigg\} \langle \bar{q}g_{s}\sigma
Gq\rangle_{\rho_N}\nonumber \\
&&-\frac{1}{48M^{2}\pi^{2}}\Bigg\{(-1+\beta)\Bigg[(1+5\beta)m_{u}
M^2-32(1+2\beta)m_ { u }p_{0}^{2}\nonumber \\
&&-4(1+\beta)m_{d}(M^2+12p_{0}^{2})\Bigg]\Bigg\}\langle\bar{q}iD_{0}iD_{0}
q\rangle_{\rho_N} \nonumber \\
&&-\frac{1}{144\pi^{2}}\Bigg\{\Bigg[3(\beta-1)m_{q}\Big(4(1+\beta)m_{d}
-(1+5\beta)m_{u}\Big)\nonumber \\
&&+16\Big(5+\beta(2+5\beta)\Big)p_{0}^{2}\Bigg]\Bigg\}\langle q^{\dag}iD_{0}
q\rangle_{\rho_N}
+\frac{1}{36\pi^{2}}\Bigg\{\Bigg[5+\beta(2+5\beta)\Bigg]m_q
p_{0}^{2}\Bigg\}\langle \bar{q}
q\rangle_{\rho_N},\nonumber \\
\end{eqnarray}
\begin{eqnarray}\label{structurep}
\hat{B}\Pi_{p}^{O}&=&\frac{1}{72 \pi^{2}}\int_{0}^{s_{0}}ds
e^{-s/M^{2}}\Bigg\{15\langle
q^{\dag}q\rangle_{\rho_N}+3\beta(2+5\beta)\langle
q^{\dag}q\rangle_{\rho_N}\Bigg\}\nonumber \\
&&+\frac{1}{576M^{2}\pi^{2}}\Bigg\{-3\Big(1+3\beta(2+\beta)\Big)M^2 
+8\Big(5+\beta(2+5\beta)\Big)p_0^{2} \Bigg\} \langle
q^{\dag}g_{s}\sigma Gq\rangle_{\rho_N}\nonumber \\
&&-\frac{1}{12M^{2}\pi^{2}}\Bigg\{\Bigg[5+\beta(2+5\beta)\Bigg](M^2-2p_{0}
^{2})\Bigg\}\langle q^{\dag}iD_{0}iD_{0}
q\rangle_{\rho_N} \nonumber \\
&&-\frac{1}{4\pi^{2}}\Bigg\{(\beta-1)m_q\Bigg[3(1+\beta)m_{d}+(2+4\beta)m_{u}
\Bigg]\Bigg\}\langle q^{\dag}
q\rangle_{\rho_N},
\end{eqnarray}
\begin{eqnarray}\label{structureu}
\hat{B}\Pi_{u}^{E}(p)&=&\frac{1}{72 \pi^{2}}\int_{0}^{s_{0}}ds
e^{-s/M^{2}}\Bigg[-3\Big(5+\beta(2+5\beta)\Big)\langle q^{\dag}g_{s}\sigma
Gq\rangle_{\rho_N}\nonumber \\
&&-9(-1+\beta)m_q\Big(3(1+\beta)m_d+2m_u(1+2\beta)
\Big)\langle q^{\dag}
q\rangle_{\rho_N}
\nonumber \\
&&+3\langle q^{\dag}
q\rangle_{\rho_N}s)\Bigg]+\frac{1}{128 \pi^{2}}\int_{0}^{s_{0}}ds 
e^{-s/M^{2}}5(1+\beta^2) \langle q^{\dag}g_{s}\sigma Gq\rangle_{\rho_N}
\nonumber \\
&&+\frac{1}{24\pi^{2}}\Bigg[5+\beta(2+5\beta)\Bigg]p_{0 } ^ { 2 }
\langle
q^{\dag}g_{s}\sigma Gq\rangle_{\rho_N} \nonumber \\
&&+\frac{1}{2\pi^{2}}\Bigg[5+\beta(2+5\beta)\Bigg]p_{0}^{2}\langle
q^{\dag}iD_{0}i D_{0}
q\rangle_{\rho_N},
\end{eqnarray}
\begin{eqnarray}\label{structureu}
\hat{B}\Pi_{u}^{O}(p)&=&\frac{1}{72 \pi^{2}}\int_{0}^{s_{0}}ds
e^{-s/M^{2}}\Bigg[
5\Big(5+\beta(2+5\beta)\Big)m_q  \langle \bar{q}
q\rangle_{\rho_N}\nonumber\\
&&+2\Big(5+\beta(2+5\beta)\Big)(-10\langle q^{\dag}iD_{0}
q\rangle_{\rho_N}\nonumber \\
&&+\frac{1}{96\pi^{2}}\Bigg\{(\beta-1)\Bigg[
8(1+\beta)m_{d}+3(3+7\beta)m_{u}
\Bigg]\Bigg\} \langle \bar{q}g_{s}\sigma
Gq\rangle_{\rho_N}\nonumber
\\
&&+\frac{1}{12\pi^{2}}\Bigg\{(\beta-1)\Bigg[8(1+\beta)m_d+3(3+7\beta)m_u\Bigg]
\Bigg\}\langle\bar{q}iD_{0}iD_{0}
q\rangle_{\rho_N}\nonumber \\
&&+\frac{1}{12\pi^{2}}\Bigg\{(\beta-1)m_q\Bigg[
4(1+\beta)m_d -(1+5\beta)m_u\Bigg]\Bigg\}\langle
q^{\dag}iD_{0}
q\rangle_{\rho_N},
\end{eqnarray}
\begin{eqnarray}\label{structures}
\Pi_{S}^{E}(p)&=&-\frac{1}{64 \pi^{4}}\int_{0}^{s_{0}}ds
e^{-s/M^{2}}s^2 \Bigg[(\beta-1)^2
m_d+6(\beta^2-1)m_u\Bigg]\nonumber \\
&&-\frac{1}{32 \pi^{2}}(\beta-1)\int_{0}^{s_{0}} ds
e^{-s/M^{2}} \Bigg\{\Big((5+7\beta)\langle
\bar{q}g_{s}\sigma
Gq\rangle_{\rho_N}\Big) \nonumber \\
&&+4m_q\Bigg[(\beta-1)m_d+6(\beta+1)m_u\Bigg]\langle \bar{q}
q\rangle_{\rho_N}-2(5+7\beta)s\langle \bar q
q\rangle_{\rho_N}\Bigg\}\nonumber \\
&&+\frac{\langle
g_{s}^{2}G^{2}\rangle_{\rho_N}}{512 \pi^{4}}(\beta-1)\int_{0}^{s_{0}}ds
e^{-s/M^{2}}\Bigg[\beta
m_{d}-6(1+\beta)m_{u}\Bigg]
\nonumber \\
&&+\frac{1}{128 \pi^{4}}(\beta-1)\beta\int_{0}^{s_{0}}ds
e^{-s/M^{2}}\langle
\bar{q}g_{s}\sigma Gq\rangle_{\rho_N} 
\nonumber \\
&&+\frac{1}{192 \pi^{2}}(\beta-1)(20+29\beta)p_{0}^{2}\langle
\bar{q}g_{s}\sigma Gq\rangle_{\rho_N} \nonumber \\
&&-\frac{1}{24 \pi^{2}}\Bigg[20+(9-29\beta)\beta 
\Bigg]p_{0}^{2}\langle \bar{q}iD_{0}iD_{0}
q\rangle_{\rho_N}\nonumber \\
&&+\frac{1}{12\pi^{2}}(\beta-1)m_q\Bigg[(\beta-1)
m_d+6(\beta+1)m_u\Bigg]p_{0}^{2}\langle \bar{q}q\rangle_{\rho_N},
\end{eqnarray}
and
\begin{eqnarray}\label{structures}
\Pi_{S}^{O}(p)&=&-\frac{1}{32 \pi^{2}}(\beta-1)\int_{0}^{s_{0}} ds
e^{-s/M^{2}} \Bigg\{4\Bigg[m_q(5+7\beta)+m_d (1  -\beta)\nonumber \\
&&-6(1+\beta)m_u\Bigg] \langle q^{\dag}
q\rangle_{\rho_N}\Bigg\}\nonumber \\
&&+\frac{1}{192 M^{2} \pi^{2}}(\beta-1)\Bigg\{3+(8+7\beta)m_u 
M^{2}+48(1+\beta)m_u p_{0}^{2}\nonumber \\
&&+4m_d\Bigg[M^{2}(1-4\beta)
+2(\beta-1)p_{0}^{2}\Bigg]\Bigg\}\langle
q^{\dag}g_{s}\sigma Gq\rangle_{\rho_N} \nonumber \\
&&+\frac{1}{4
M^{2}\pi^{2}}(\beta-1)\Bigg[(\beta-1)m_d+6(\beta+1)m_u\Bigg](M^{
2}+2p_{0}^{2})\langle q^{\dag}iD_{0}iD_{0}q\rangle_{\rho_N} \nonumber \\
&&-\frac{1}{24\pi^{2}}(\beta-1)\Bigg[\beta m_q-8m_d(1-\beta)
+48(1+\beta)m_u\Bigg]\langle q^{\dag}iD_{0}q\rangle_{\rho_N}
\end{eqnarray}
where $s_0$ is the continuum threshold.
Having calculated both the hadronic and OPE sides of the correlation function, now, we equate these two sides for all structures to find the corresponding QCD sum rules. For instance, in the case of the 
structure $ \!\not\!{p}$, we have
\begin{eqnarray}
-\lambda_N^{*2}e^{-\mu^2/M^2}=\hat{B}\Pi_{p}^{OPE}.
\end{eqnarray}
To find the  mass sum rule, we eliminate the $\lambda_N^{*2}$ in the above equation, as a result of which we get
\begin{eqnarray} \label{ratio}
\mu^2=\frac{\frac{\partial}{\partial(-\frac{1}{M^2})}\Big(\hat{B}\Pi_{p}^{OPE}\Big)}{\hat{B}\Pi_{p}^{OPE}}.
\end{eqnarray}
\begin{table}[ht!]
\centering
\rowcolors{1}{lightgray}{white}
\begin{tabular}{cc}
\hline \hline
   Input Parameters  &  Values    
           \\
\hline \hline
$p_0   $          &  $1  $ $GeV$      \\
$ m_{u}   $          &  $2.3  $ $MeV$       \\
$ m_{d}   $          &  $4.8  $ $MeV$       \\
$ \rho_{N}     $          &  $(0.11)^3  $ $GeV^3$        \\
$ \langle q^{\dag}q\rangle_{\rho_N}    $          &  $\frac{3}{2}\rho_{N}$         \\
$ \langle\bar{q}q\rangle_{0}           $          &  $ (-0.241)^3    $ $GeV^3$          \\
$ m_{q}      $          &  $0.5(m_{u}+m_{d})$                 \\
$ \sigma_{N}            $          &  $0.045 ~  $GeV$ $                  \\
$  \langle\bar{q}q\rangle_{\rho_N}  $          &  $ \langle\bar{q}q\rangle_{0}+\frac{\sigma_{N}}{2m_
{q}}
\rho_{N}$                  \\
$  \langle q^{\dag}g_{s}\sigma
Gq\rangle_{\rho_N}  $          &  $ -0.33~GeV^2 \rho_{N}$                  \\
$  \langle q^{\dag}iD_{0}q\rangle_{\rho_N}  $          &  $0.18 ~GeV \rho_{N}$                  \\
$  \langle\bar{q}iD_{0}q\rangle_{\rho_N}  $          &  $\frac{3}{2} m_q \rho_{N}\simeq0 $                  \\
$  m_{0}^{2}  $          &  $ 0.8~GeV^2$                  \\
$   \langle\bar{q}g_{s}\sigma Gq\rangle_{0} $          &  $m_{0}^{2}\langle\bar{q}q\rangle_{0} $                  \\
$  \langle\bar{q}g_{s}\sigma
Gq\rangle_{\rho_N}  $          &  $\langle\bar{q}g_{s}\sigma Gq\rangle_{0}+3~GeV^2\rho_{N} $                  \\
$ \langle  \bar{q}iD_{0}iD_{0}q\rangle_{\rho_N} $          &  $ 0.3~GeV^2\rho_{N}-\frac{1}{8}\langle\bar{q}g_{s}
\sigma
Gq\rangle_{\rho_N}$                  \\
$  \langle
q^{\dag}iD_{0}iD_{0}q\rangle_{\rho_N}  $          &  $0.031~GeV^2\rho_{N}-\frac{1}{12}\langle
q^{\dag}g_{s}
\sigma Gq\rangle_{\rho_N} $                  \\
$\langle \frac{\alpha_s}{\pi} G^{2}\rangle_{0}$ & $(0.33\pm0.04)^4~GeV^4$\\
$\langle \frac{\alpha_s}{\pi} G^{2}\rangle_{\rho_N}$ & $\langle \frac{\alpha_s}{\pi} G^{2}\rangle_{0}-0.65~GeV \rho_N$\\
 \hline \hline
\end{tabular}
\caption{Numerical values for input parameters \cite{Cohen,Nielsen,XJ1,Cohen45}. The value presented for $\rho_N$ corresponds to the nuclear matter saturation density which is used in numerical analysis. }
\end{table}
 
 \section{Numerical results and discussion}

This section is devoted to the numerical analysis of the sum rules for the mass and residue obtained in the previous section at nuclear matter.
 We discuss how the results in dense medium deviate from those obtained via vacuum sum rules. For this aim, we need the numerical values of
the quark masses as well as the in-medium quark-quark, quark-gluon and gluon-gluon condensates that are calculated 
in 
\cite{Cohen,Nielsen,XJ1,Cohen45}. Each condensate at dense nuclear medium ($\langle {\cal O}\rangle_{\rho_N}$) is written in terms of its vacuum values ($\langle  {\cal O}\rangle_0$) 
and its value between one-nucleon states
 ($\langle {\cal O}\rangle_N$) at the low nuclear density limit, i.e.
$\langle {\cal O}\rangle_{\rho_N}=\langle 0| {\cal O}|0\rangle+\rho_N\langle N| {\cal O}|N\rangle
=\langle  {\cal O}\rangle_0+\rho_N\langle {\cal O}\rangle_N$. 
We collect the numerical values of the input parameters in Table 1.

Looking at the sum rules for the physical quantities under consideration we see that they include three auxiliary parameters, namely, continuum threshold $s_0$,
Borel mass parameter $M^2$ and mixing parameter $\beta$ that should be fixed at this point. The standard criteria in QCD sum rules demand that the physical quantities show good stability with respect to these quantities
at their working regions. As the mass sum rule is the ratio of two 
sum rules (see Eq. \ref{ratio}) including these auxiliary parameters, it may not lead to a reliable region. For this reason, we use the sum rule for the residue to find the reliable regions for the helping parameters.
The working region for the Borel mass parameter is found as follows. 
The upper bound on this parameter is found by demanding that  the contributions of the higher states and continuum are sufficiently suppressed and the ground state constitutes a large part of
 the whole dispersion integral, i.e.
\bea
\label{nolabel}
{\ds \int_{0}^{s_0}\ds  \rho(s) e^{-s/M^2} \over
\ds \int_{0}^\infty \rho(s) e^{-s/M^2}} ~~>~~ 1/2, 
\eea
should be satisfied. The lower bound on $M^2$ is calculated requiring that the perturbative part exceeds the non-perturbative one and the contributions of the operators with higher dimensions are small, i.e. 
the OPE converges. 
These requirements  lead to the interval $0.8~GeV^2\leqslant M^2 \leqslant 1.2~GeV^2$ for the Borel mass squared. The continuum threshold is not totally arbitrary but it depends on the 
energy of the first excited state with the same quantum numbers as the interpolating current. We choose the interval $s_0=(1.5-2.0)~GeV^2$ for this parameter. Our numerical calculations depict 
that, in this interval, the physical quantities depend weakly on this parameter  and the results show good stability with respect to the variations of Borel mass parameter in its working region.
\begin{figure}[h!]
\centering
\begin{tabular}{ccc}
\includegraphics[totalheight=6cm,width=7cm]{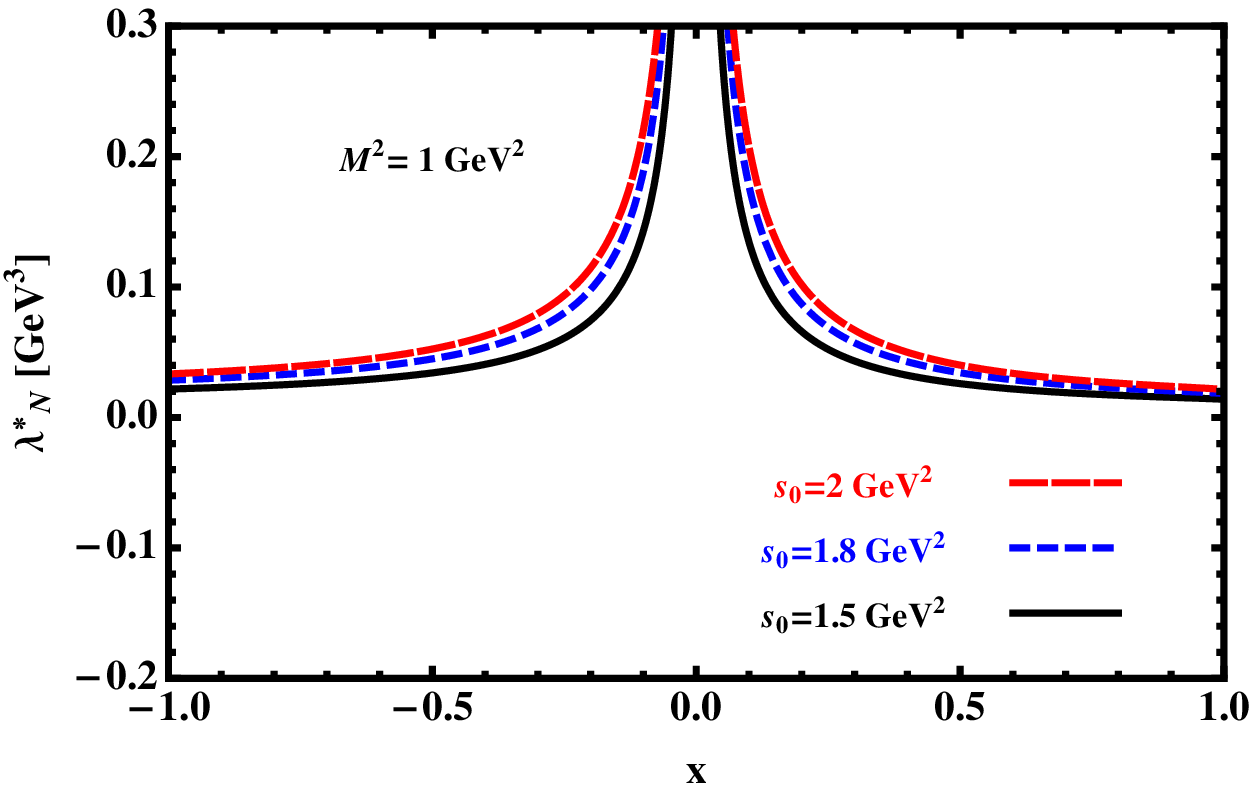}
\includegraphics[totalheight=6cm,width=7cm]{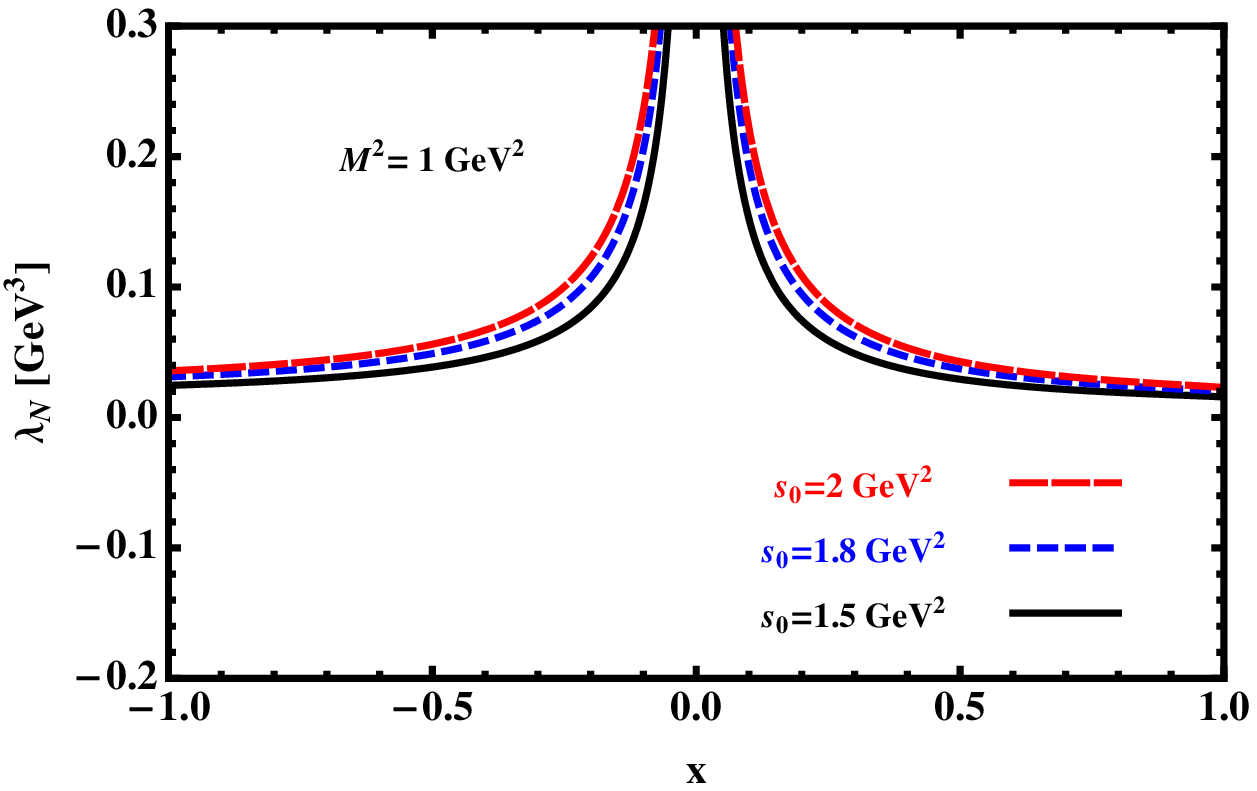}
\end{tabular}
\caption{The residue in nuclear matter versus $x$   (left panel). The residue in vacuum versus $x$  (right panel).}
\end{figure}
\begin{figure}[h]
\centering
\begin{tabular}{cc}
\includegraphics[totalheight=6cm,width=7cm]{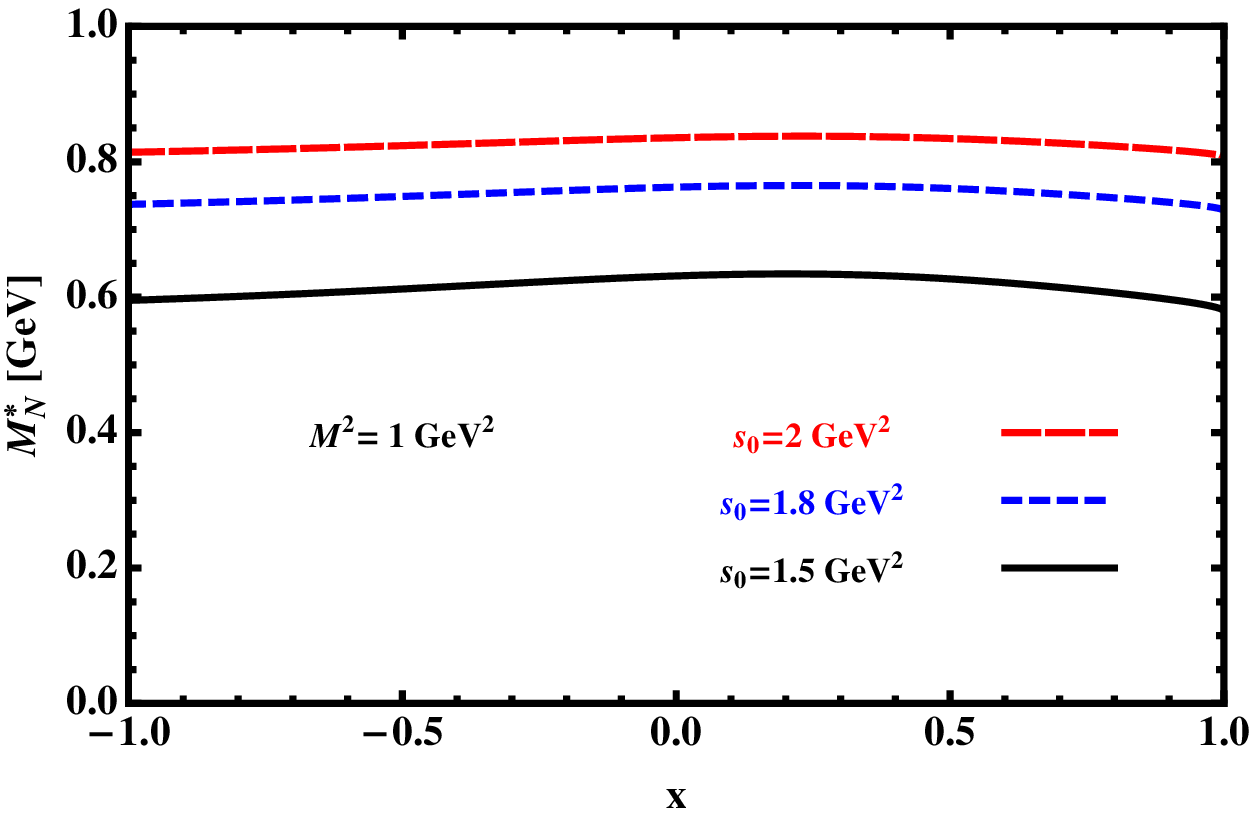}
\includegraphics[totalheight=6cm,width=7cm]{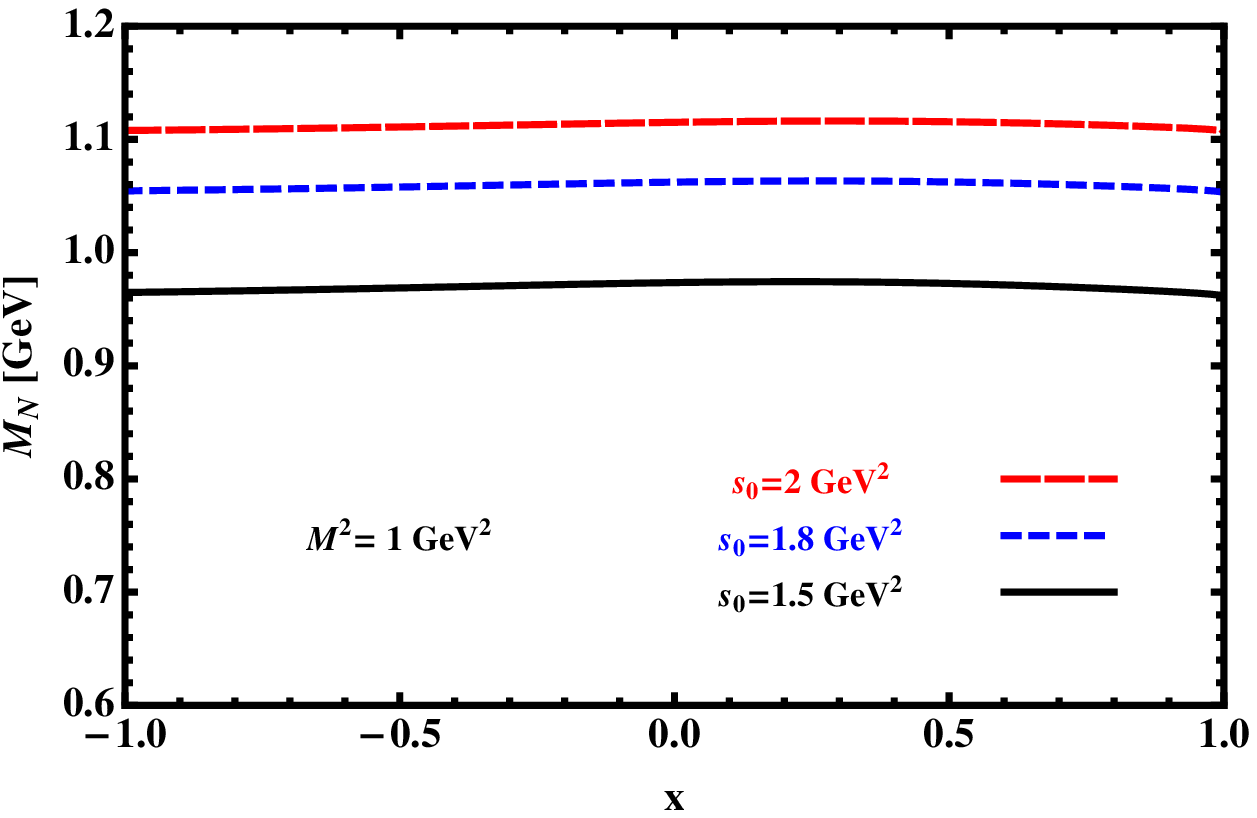}
\end{tabular}
\caption{The nucleon mass in nuclear matter versus $x$ (left panel). The nucleon mass in vacuum
versus $x$ (right panel).  }
\end{figure}
\begin{figure}[h]
\centering
\begin{tabular}{ccc}
\includegraphics[totalheight=6cm,width=7cm]{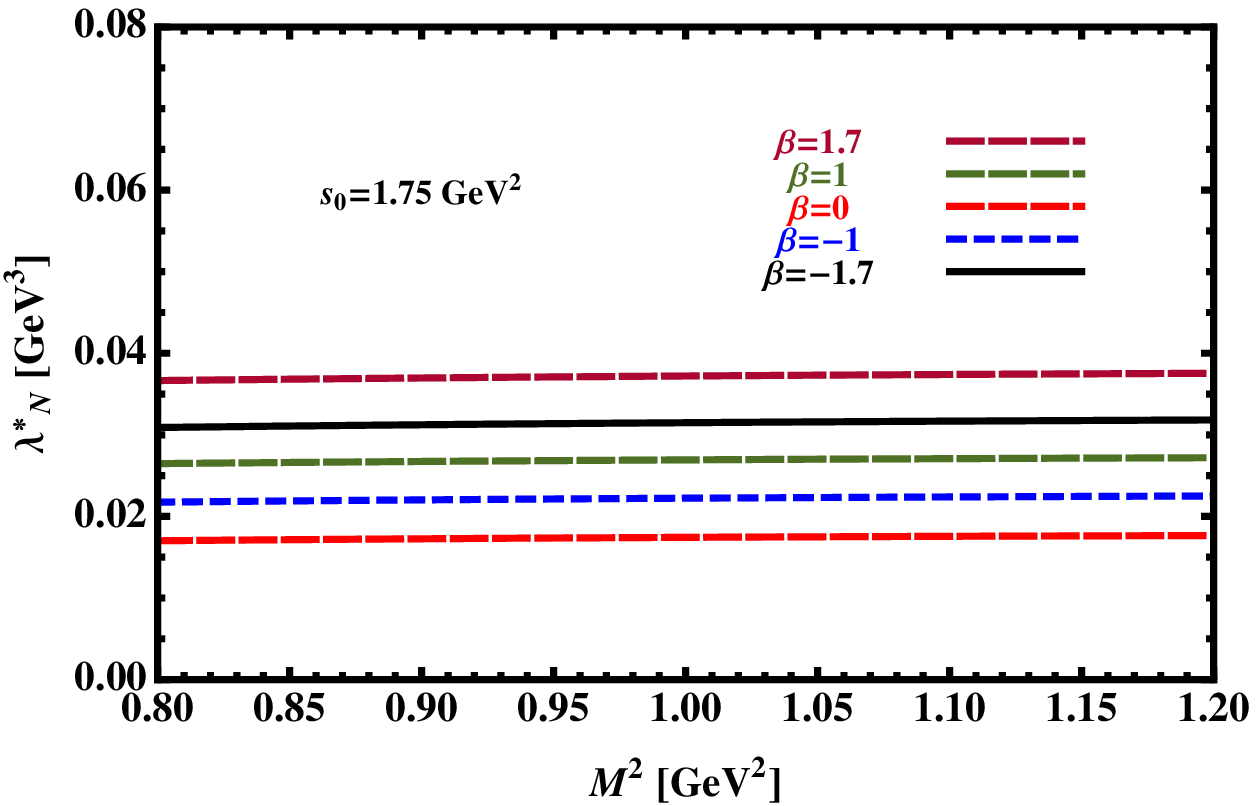}
\includegraphics[totalheight=6cm,width=7cm]{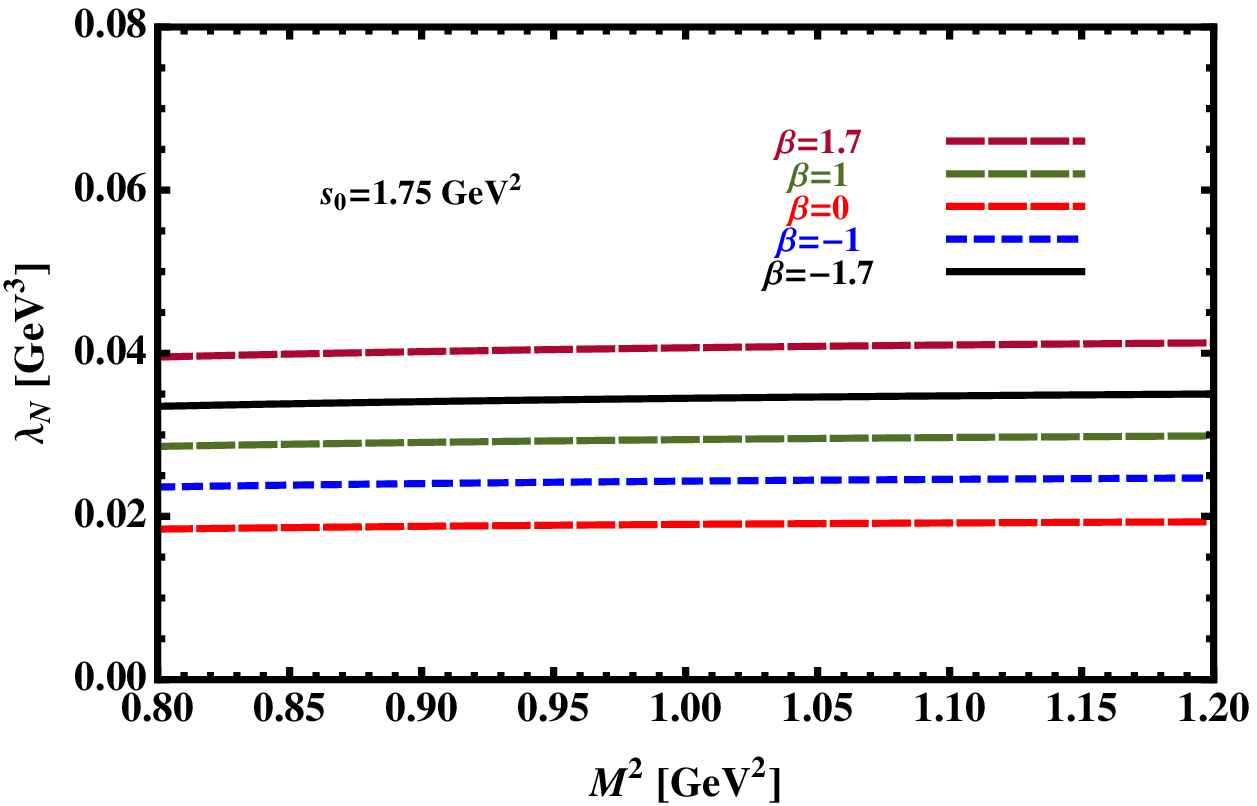}
\end{tabular}
\caption{The residue in nuclear matter versus Borel mass $M^{2}$  (left panel).The residue in vacuum versus
Borel mass $M^{2}$  (right panel).}
\end{figure}  
\begin{figure}[h]
\centering
\begin{tabular}{ccc}
\includegraphics[totalheight=6cm,width=7cm]{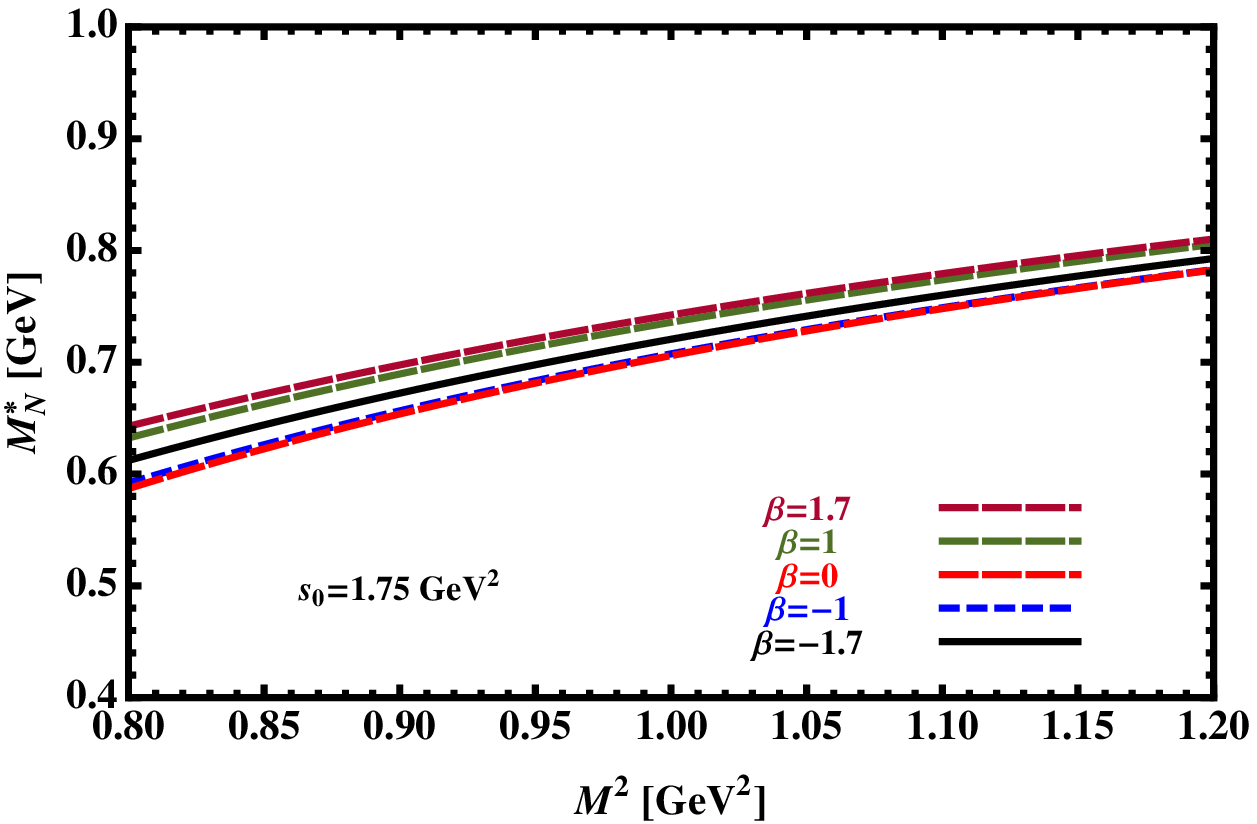}
\includegraphics[totalheight=6cm,width=7cm]{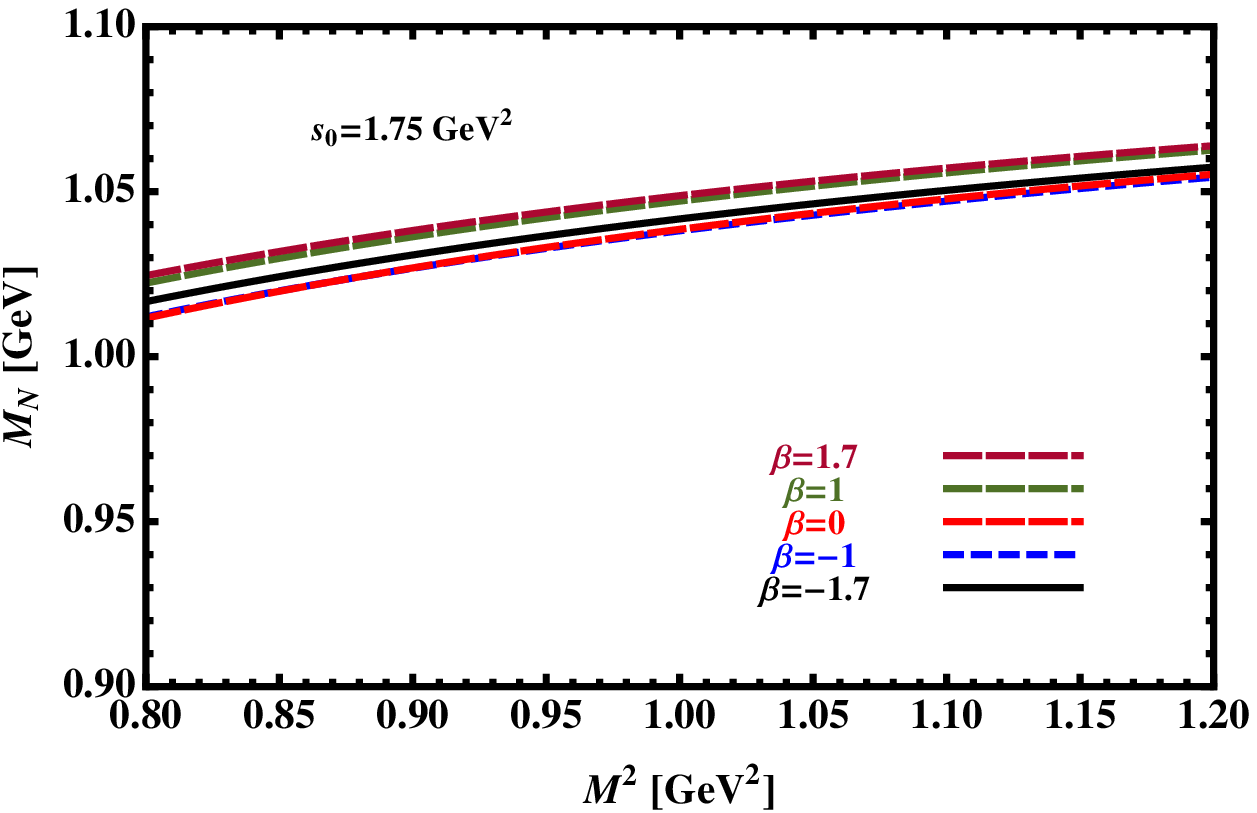}
\end{tabular}
\caption{The nucleon mass in nuclear matter versus Borel mass $M^{2}$  (left panel). The nucleon mass in vacuum versus
Borel mass $M^{2}$  (right panel).}
\end{figure}

Finally, the physical quantities under consideration should be independent of the mixing parameter $\beta$. To find the working region for this parameter, we look at the variation of the residue of
 the nucleon  with respect to this parameter. To better cover the whole   range 
 $-\infty \leqslant \beta\leqslant \infty$, can be mathematically  taken by this parameter, we plot the the residue  with respect to $x=cos\theta$, where $\beta=tan\theta$ at fixed values of the continuum 
threshold and Borel mass parameter
 for both nuclear medium and vacuum in figure 1.
From this figure, we see that in the intervals $-1\leqslant x \leqslant -0.5$ and $0.5\leqslant x \leqslant 1$ the residues $\lambda^*_N$ and $\lambda_N$ are practically independent of this parameter. 
Moreover, the results of residues depend
on continuum threshold very weakly in these intervals. Note that the Ioffe current corresponding to $x\approx-0.71$ is included by these intervals.
Here, we should mention that, as also we said before, since the mass sum rule is the ratio of two sum rules, the unstable points of two sum rules in nominator and denominator coincide and cancel each other.
Such that, the masses  in nuclear matter and vacuum show roughly good stabilities with respect to $x$ in the whole  $-1\leqslant x \leqslant 1$ region (see figure 2).

Having calculated the working regions, now, we discuss the variations of the masses and residues both in nuclear matter and vacuum with respect to the variations of the auxiliary parameters and look for the shifts
in these parameters due to the nuclear medium by comparison of the results obtained in the nuclear matter as well as the vacuum. For this aim, in figures 3 and 4, we depict the variations of the residues and 
masses in the presence of nuclear matter and vacuum with respect to the Borel mass parameter at different fixed values of the $\beta$ and $s_0$ picked from their working regions. These figures also indicate that 
the physical quantities under consideration vary weakly  with respect to the helping parameters in their working regions.

\begin{table}[htdp]
\begin{center}
\rowcolors{1}{lightgray}{white}
\begin{tabular}{lcccc}
\hline \hline
& $m_{N}^{*}$ (GeV) & $m_N$ (GeV) & $ \lambda_{N}^{*2}$ ($GeV^6$) & $ \lambda_N^{2}$ ($GeV^6$)  \\
\hline \hline
Present work & $0.723\pm0.122$ & $1.045\pm$0.076 & $0.0009\pm0.0004$ & $0.0011\pm0.0005$  \\
 \cite{Ioffe}& - & 0.985 & - & $0.0012\pm 0.0006$ \\
\cite{nasrallah}& - & $0.990\pm0.050$ & - & - \\
\hline \hline
\end{tabular} \caption{Average values of the  masses and residues squared in  nuclear matter and vacuum obtained from sum rules analysis and the comparison of the results with the existing results 
of the vacuum sum rules for the Ioffe current \cite{Ioffe}, and value of the mass obtained via Ioffe current in vacuum cosidering the strangness content of the nucleon \cite{nasrallah}.}
\end{center}
\label{result}
\end{table}
\begin{figure}[h]
\centering
\begin{tabular}{ccc}
\includegraphics[totalheight=6cm,width=7cm]{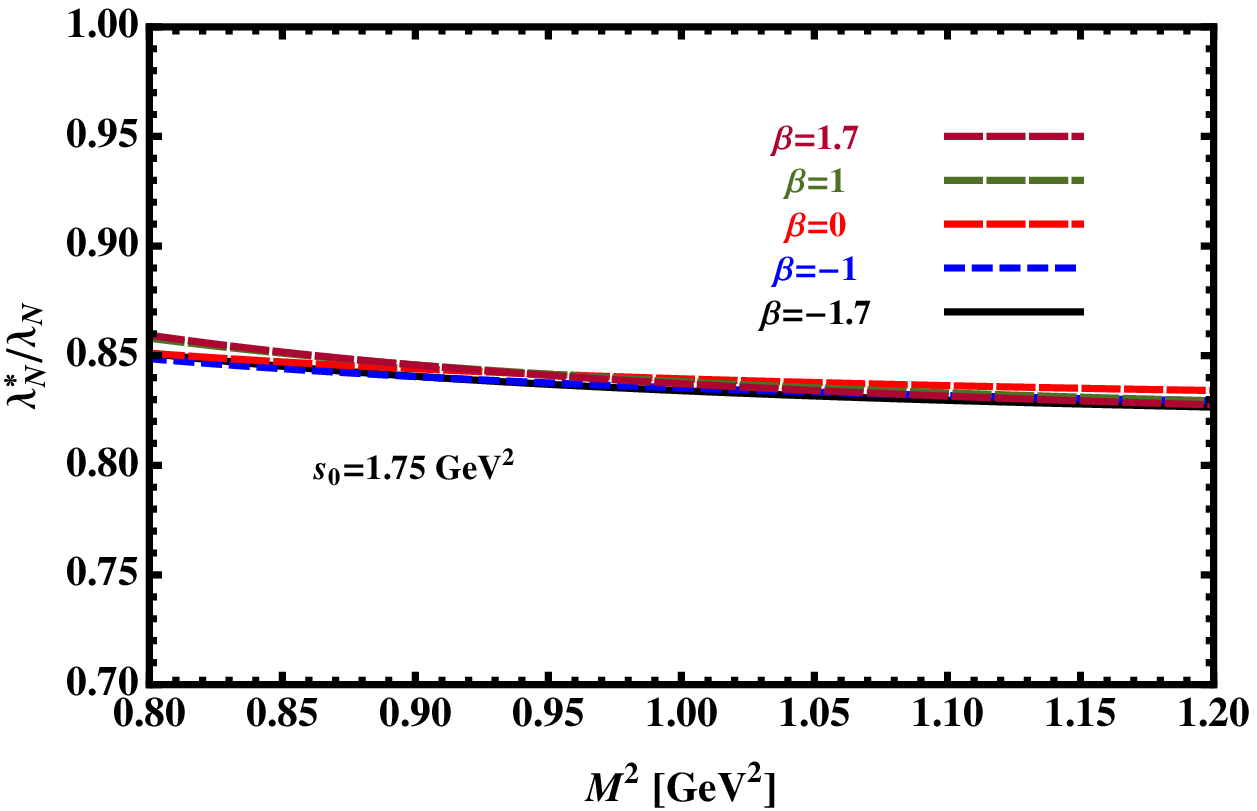}
\includegraphics[totalheight=6cm,width=7cm]{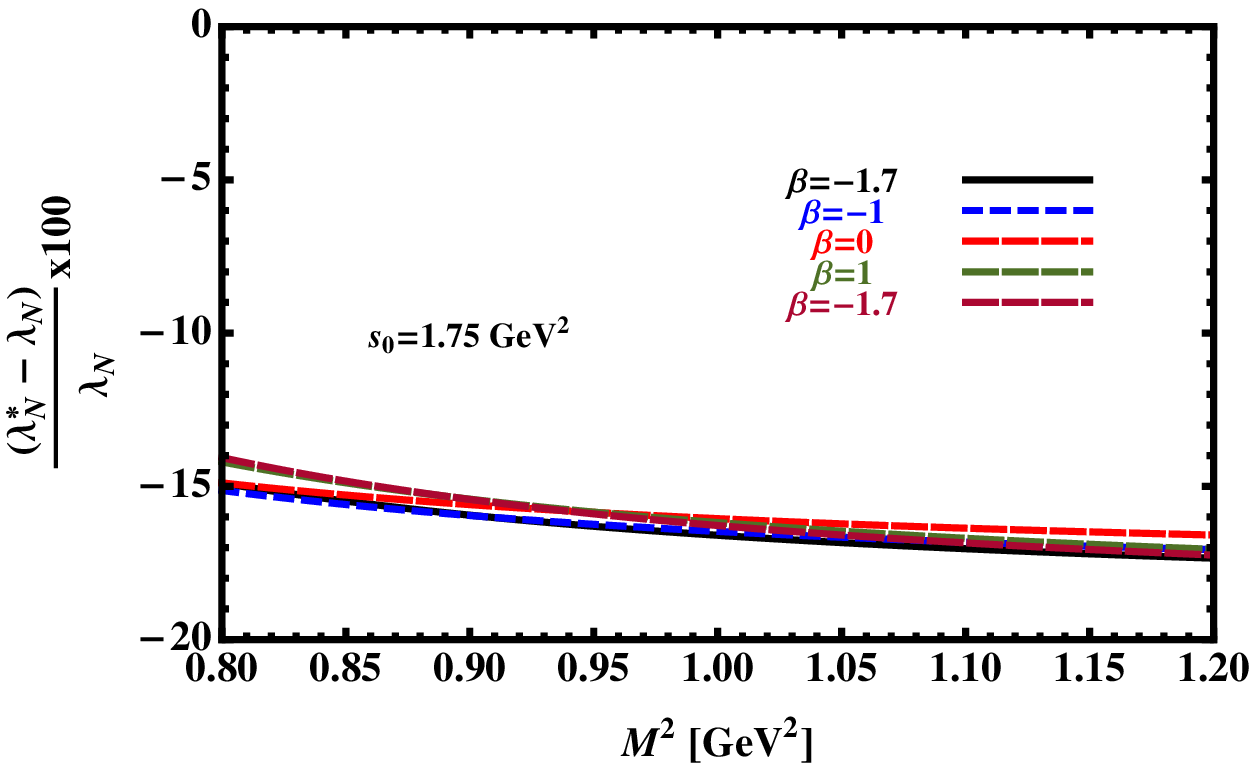}
\end{tabular}
\caption{ $\lambda_{N}^{*}/\lambda_{N}$ versus Borel mass parameter
$M^{2}$  (left panel). The percentage of the shift in the residue of the nucleon in nuclear matter compared to its vacuum value (right panel).}
\end{figure}  
\begin{figure}[h]
\centering
\begin{tabular}{ccc}
\includegraphics[totalheight=6cm,width=7cm]{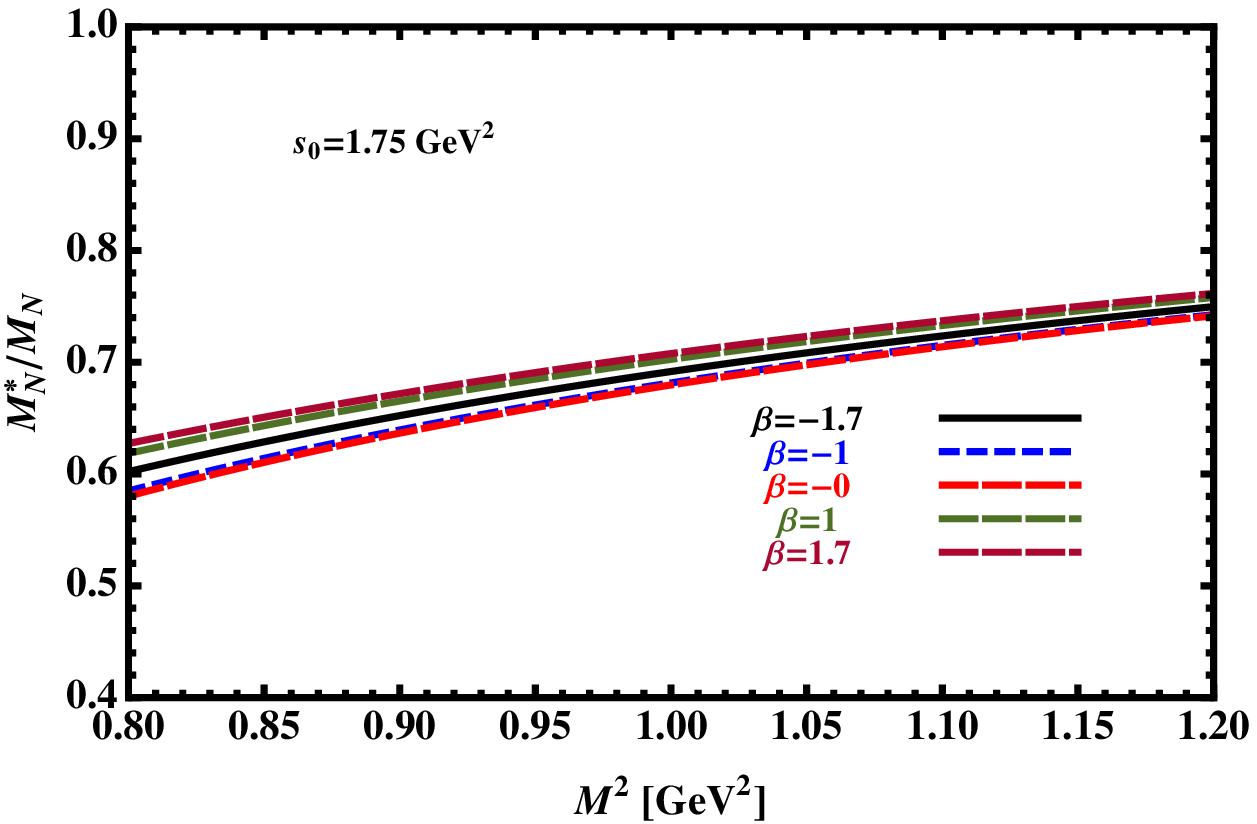}
\includegraphics[totalheight=6cm,width=7cm]{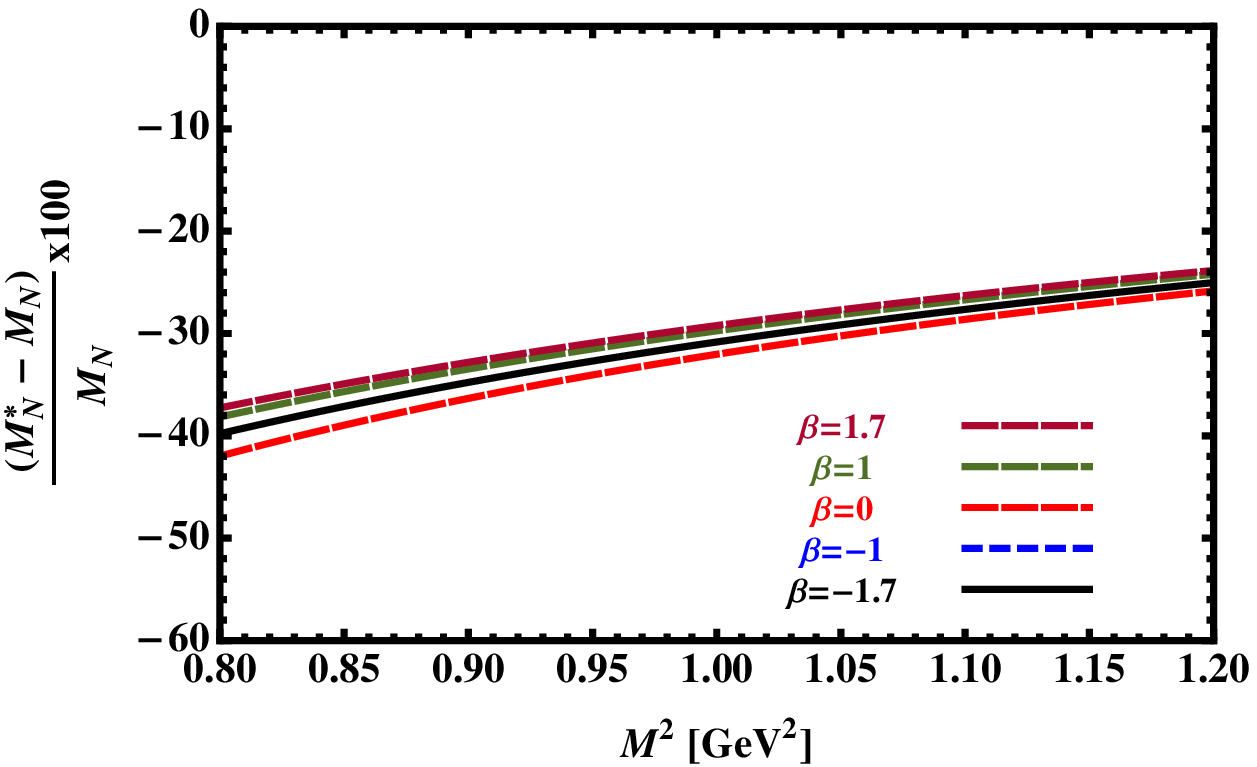}
\end{tabular}
\caption{ $m_{N}^{*}/m_{N}$ versus Borel mass parameter
$M^{2}$  (left panel). The percentage of the shift in the mass of the nucleon in nuclear matter compared to its vacuum value (right panel).}
\end{figure}       
Obtained from figures 3 and 4, we depict the average values of the residues squared and masses of the nucleon both for the nuclear medium and vacuum in Table 2 and  compare our results with the existing results 
obtained via the Ioffe current using the vacuum sum rules in this table. Note that in this table $m_N$ and $\lambda_N$ are the mass and residue of the nucleon in vacuum and are respectively  obtained from
 $m^*_N$ and $\lambda^*_N$ when $\rho_N=0$ is set.
From this table, we conclude that the average values for the residue squared and mass when $\rho_N\rightarrow0$ are consistent with the values obtained using the Ioffe current and vacuum sum rules \cite{Ioffe,nasrallah} within the errors.
We also see that the average
values of those quantities in nuclear matter show considerable shifts compared to the vacuum results. To see better how the results of the residue and mass in nuclear matter deviate from those of the vacuum,
 we depict the variations of the ratios of $\lambda_{N}^{*}/\lambda_{N}$ and $m_{N}^{*}/m_{N}$ as well as the percentages of the shifts with respect to the Borel mass squared in figures 5 and 6 at different fixed values
of the  parameter $\beta$ and the continuum threshold $s_0$.
With a quick glance at these figures, we observe that the mass and residue of the nucleon show considerable shifts from their vacuum values and the shifts are negative. In the case of the residue,  the shift
 grows roughly increasing the value of the Borel mass parameter. However, in the case of mass, the shift deceases considerably when we increase the value of the Borel mass parameter in its working region.
 Our numerical
results show that, in average, the values of the residue and mass decrease about $15\%$ and  $32\%$,  respectively compared to their values in vacuum. Note that we have used 
 the nuclear matter saturation density,  $\rho_N^{sat}=(0.11)^3$ $GeV^3$,  in our numerical analysis as well as the density dependence of some condensates in leading order (see Table 1). To check whether the results
 depend linearly on the nuclear matter density or not, we depict the dependences of, for instance,  $m_{N}^{*}/m_{N}$ and $\lambda_{N}^{*}/\lambda_{N}$ on $\rho_N/\rho_N^{sat}$ in figure 7 for fixed values of $s_0$, $M^2$
and $\beta$. From this figure we see that $\lambda_{N}^{*}/\lambda_{N}$ is  exactly linear and $m_{N}^{*}/m_{N}$ is  roughly linear in terms of  $\rho_N/\rho_N^{sat}$ and they considerably decrease increasing 
the value of the nuclear matter density.
\begin{figure}[h]
\centering
\begin{tabular}{ccc}
\includegraphics[totalheight=6cm,width=7cm]{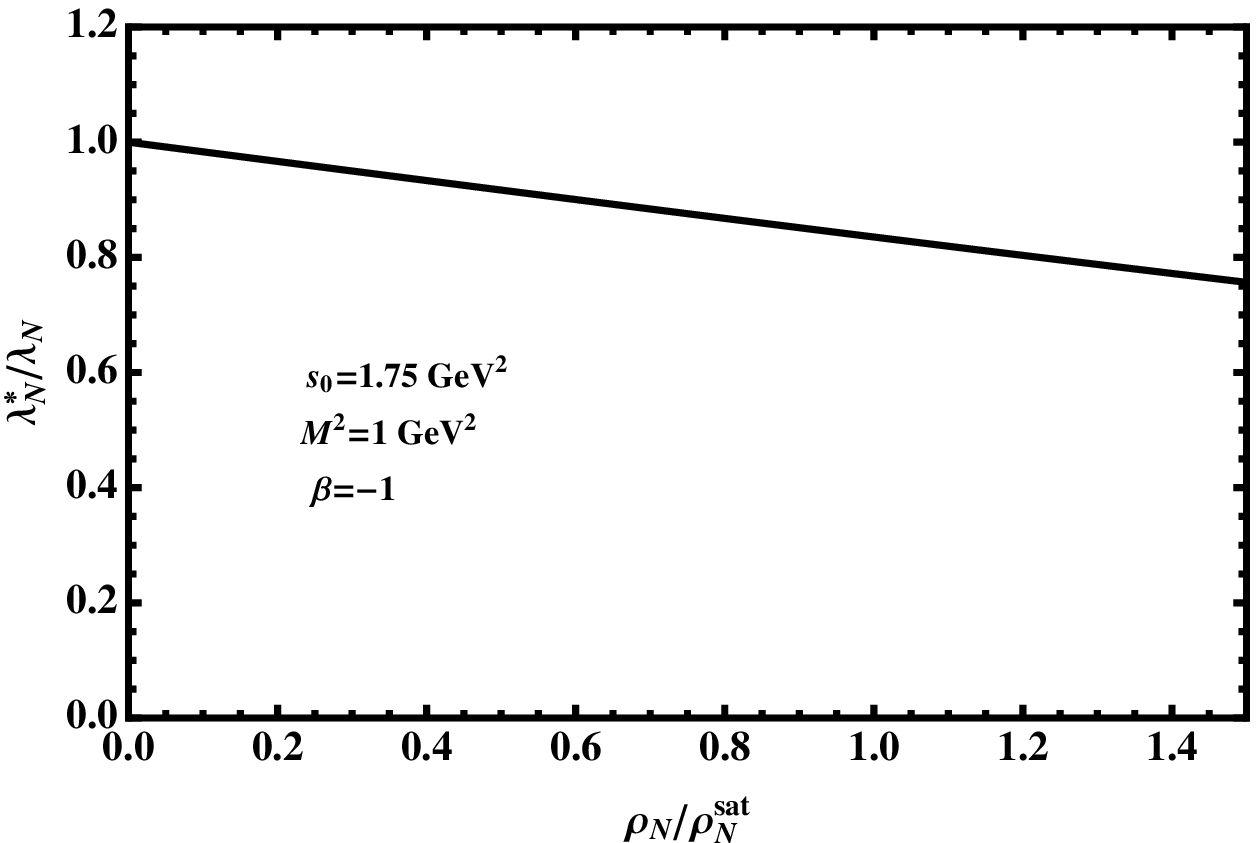}
\includegraphics[totalheight=6cm,width=7cm]{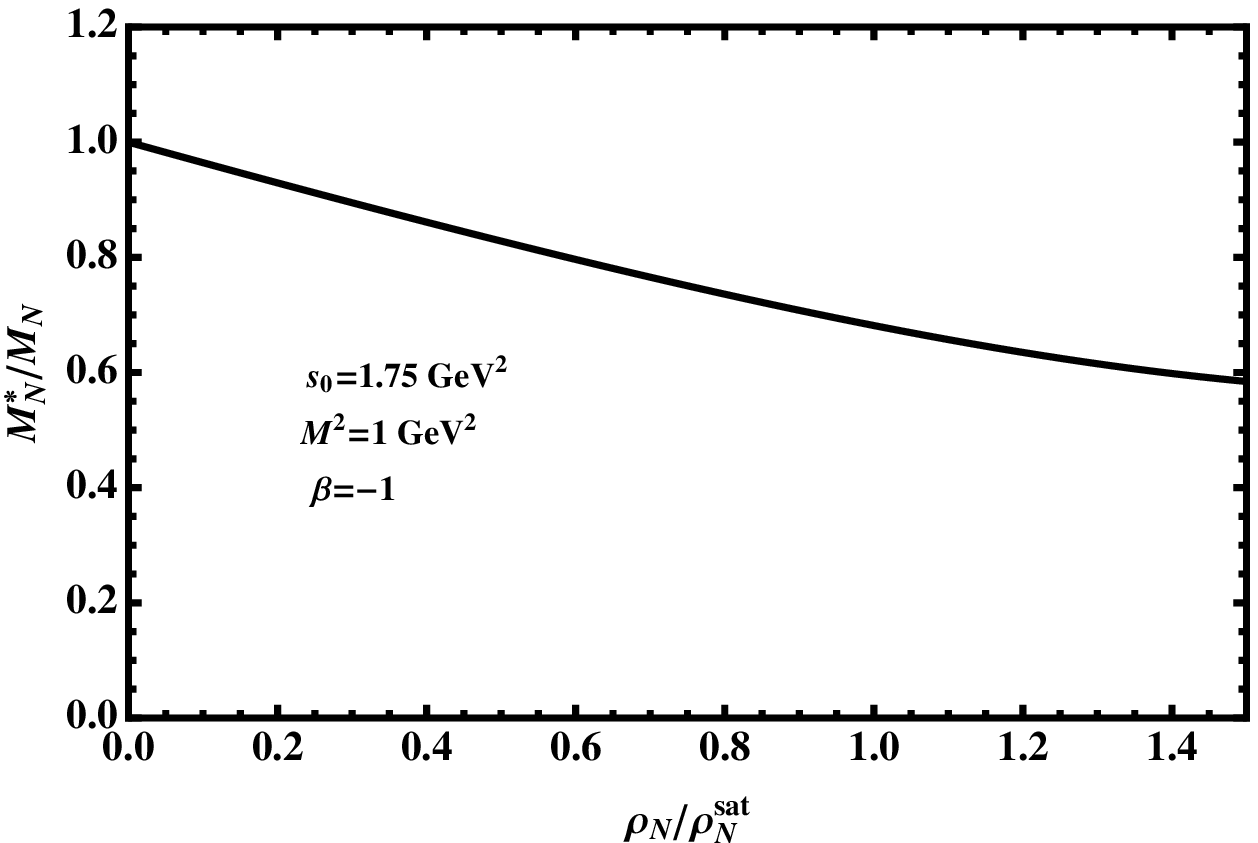}
\end{tabular}
\caption{ $\lambda_{N}^{*}/\lambda_{N}$ versus $\rho_N/\rho_N^{sat}$  (left panel). $m_{N}^{*}/m_{N}$ versus $\rho_N/\rho_N^{sat}$   (right panel).}
\end{figure} 

At the end of this section, we would like to extract the values of the  vector and scalar self-energies of the nucleon in nuclear matter. From our analysis we obtain the values $\Sigma_S=-(322\pm51)MeV$ 
and $\Sigma_0=(420\pm65)MeV$ for the scalar and time-like vector self-energies of the nucleon in nuclear medium, respectively. When we compare these results with the
 ones obtained using a model independent study in \cite{plohl1,plohl2}, i.e., 
$\Sigma_S=-(400-450)MeV$  and $\Sigma_0=(350-400)MeV$, we see that our result on time-like vector self-energies of the nucleon is consistent with the predictions of \cite{plohl1,plohl2} within the errors. 
In the case of the  scalar self-energy, although our result is  consistent with those of  \cite{plohl1,plohl2} in sign, its absolute value is smaller than those of \cite{plohl1,plohl2}.  

\section{Conclusion}
In the present work, we studied some properties of nucleon in the nuclear matter using the QCD sum rules. In particular, we calculated the mass and residue of the nucleon in nuclear medium and
looked for the shifts of the results compared to their vacuum values. Using the   interpolating current of the nucleon with an arbitrary mixing parameter, we extended the previous works on the mass of the nucleon discussed
 in the body text which  mainly use the Ioffe current. We also extended the recent study \cite{Mallik} on the residue of the nucleon pole, 
which uses  a special current corresponding to an axial-vector diquark coupled to a quark, by introducing the arbitrary mixing parameter into the interpolating current. We found the working regions for the 
three main auxiliary parameters entering the sum rules using
the obtained QCD  sum rule for the residue. Using the obtained working regions for the continuum threshold, Borel mass parameter and the mixing parameter $\beta$ entering  the  interpolating current,
 we depicted the variations of the physical quantities under consideration  with respect to the variations of the auxiliary parameters. We observed considerable negative shifts in the values of
 the mass and residue of
the nucleon in nuclear matter compared to their values in vacuum. The results of the residue  and mass reduce about $15\%$ and  $32\%$,  respectively due to the nuclear medium.
We also extracted the values of the scalar and time-like vector self-energies of the nucleon in nuclear medium and compared the obtained results with the predictions of  model independent studies \cite{plohl1,plohl2}.

 The obtained results for
the mass and residue in nuclear matter can  be used in theoretical determinations of the electromagnetic properties of the nucleon and its strong couplings to other hadrons in  nuclear medium.


\end{document}